%% file: main.tex
\documentclass[runningheads]{class/llncs}

\usepackage{class/accv}

\newcommand{\datasetabbrev}{Guide3D} 

\usepackage{import}
\usepackage{class/tudor}
\usepackage{class/accvabbrv}

\usepackage[accsupp]{axessibility}  

\usepackage[bookmarks=false,breaklinks,colorlinks,citecolor=accvblue,pdfstartview={FitH}]{hyperref}

\usepackage{orcidlink}

\begin{document}

\title{\datasetabbrev{}: A Bi-planar X-ray Dataset for \\ 3D Shape Reconstruction} 

\titlerunning{\datasetabbrev{}}

\author{
Tudor Jianu\inst{1}\orcidlink{0000-0003-0324-0950} \and
Baoru Huang\inst{2}\orcidlink{0000-0002-4421-652X} \and
Hoan Nguyen\inst{3}\orcidlink{0009-0003-9605-2939} \and
Binod Bhattarai\inst{4}\orcidlink{Binod Bhattarai} \and \\
Tuong Do\inst{1}\orcidlink{0000-0002-3290-3787} \and
Erman Tjiputra\inst{5}\orcidlink{0009-0003-6909-4623} \and
Quang Tran\inst{6}\orcidlink{0000-0001-5839-5875} \and
Pierre Berthet-Rayne\inst{1}\orcidlink{0000-0001-9118-4877} \and    
Ngan Le\inst{7}\orcidlink{0000-0003-2571-0511} \and
Sebastiano Fichera\inst{1}\orcidlink{0000-0003-1006-4959} \and
Anh Nguyen\inst{1}\orcidlink{0000-0002-1449-211X}
}



\authorrunning{T.~Jianu~\etal}

\institute{
University of Liverpool, Liverpool, United Kingdom \email{\{t.jianu,sebastiano.fichera,anh.nguyen\}@liverpool.ac.uk} \and
Imperial College London \and 
University of Information Technology - VNUHCM \and 
University of Aberdeen \and 
AIOZ Singapore \and 
University of Arkansas 
}

\maketitle

\begin{abstract}
    Endovascular surgical tool reconstruction represents an important factor in advancing endovascular tool navigation, which is an important step in endovascular surgery. However, the lack of publicly available datasets significantly restricts the development and validation of novel machine learning approaches. Moreover, due to the need for specialized equipment such as biplanar scanners, most of the previous research employs monoplanar fluoroscopic technologies, hence only capturing the data from a single view and significantly limiting the reconstruction accuracy. To bridge this gap, we introduce \datasetabbrev{}, a bi-planar X-ray dataset for 3D reconstruction. The dataset represents a collection of high resolution bi-planar, manually annotated fluoroscopic videos, captured in real-world settings. Validating our dataset within a simulated environment reflective of clinical settings confirms its applicability for real-world applications. Furthermore, we propose a new benchmark for guidewrite shape prediction, serving as a strong baseline for future work.  \datasetabbrev{} not only addresses an essential need by offering a platform for advancing segmentation and 3D reconstruction techniques but also aids the development of more accurate and efficient endovascular surgery interventions. Our project is available at \href{https://airvlab.github.io/guide3d/}{https://airvlab.github.io/guide3d/}.
  \keywords{Endovascular Dataset \and 3D Reconstruction}
\end{abstract}

\input{sections/0_introduction}
\input{sections/1_related_works}
\input{sections/2_the_dataset}
\input{sections/3_shape_prediction}
\input{sections/4_experiments}
\input{sections/5_conclusion}

\bibliographystyle{splncs04}
\bibliography{main}
\end{document}

%% file: sections/0_introduction.tex
\section{Introduction}~\label{sec:introduction}
Minimally invasive surgery has revolutionized endovascular intervention, offering less intrusive options with expedited recovery periods~\cite{puschel2022robot}. These procedures' success relies on the precise navigation and manipulation of instruments such as \textit{guidewires} and \textit{catheters}. The procedure primarily relies on 2D visualization methods for guidance, with \textit{monoplanar fluoroscopy} being the most prominent due to its minimal interference with surgical workflows and reasonable financial cost setup~\cite{rafii2014current}. However, despite their widespread use, these conventional imaging techniques present significant limitations. Among these, one primary issue is depth perception, which poses challenges in accurately visualizing surgical instruments~\cite{hoffmann2013reconstruction, hoffmann2015electrophysiology,huang2022simultaneous,huang2022self}. This limitation increases the risk of excessive contact between the surgical instruments and arterial walls, potentially endangering patient safety and diminishing the procedure's efficacy.

In endovascular intervention, depth perception crucially relies on multi-view imaging systems, such as biplanar scanners, which enable shape reconstruction by integrating multiple angular views~\cite{burgner2011toward,wagner20164d,hoffmann2015electrophysiology,ambrosini2017fully} and employing epipolar geometry based reconstruction~\cite{baur2016automatic}. However, two main challenges impede the wider adoption and efficacy of these systems~\cite{ramadani2022survey}: \textit{i)} the difficulty of accurately segmenting images for successful shape reconstruction, increased by the lack of datasets to evaluate segmentation methods, and \textit{ii)} the requirement for specialized biplanar scanners, which are not widely available in clinical settings due to expensive financial cost~\cite{burgner2011toward}. These issues highlight the \textit{critical need for comprehensive datasets} that can improve segmentation algorithm precision and guidewire reconstruction methodologies, accelerating the development of more versatile imaging technologies.

\begin{figure}[t]
    \centering
    \includegraphics[width=\textwidth]{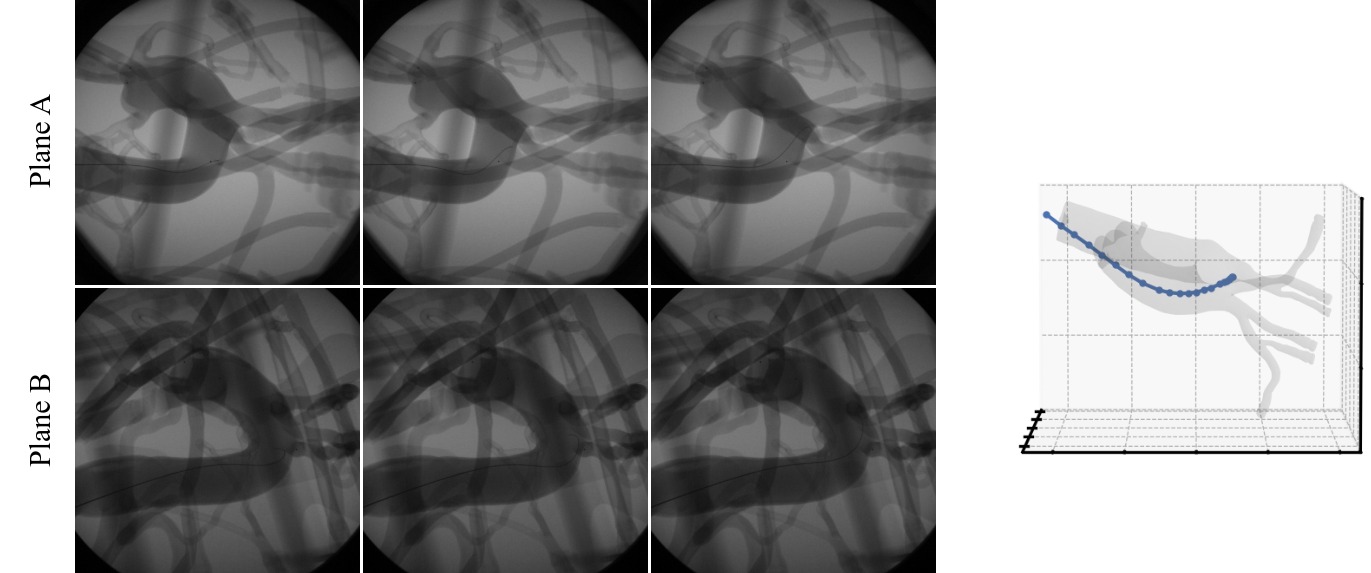}
    \caption{\textbf{Dataset Overview:} \datasetabbrev{} contains 8,746 manually annotated frames from two views for 3D reconstruction (left), from which the reconstruction is derived (right).}
    \label{fig:overview}
    \vspace{-4ex}
\end{figure}

In this paper, we introduce \textbf{\datasetabbrev{}}, a dataset tailored for advancing \textit{3D reconstruction} within endovascular navigation. \datasetabbrev{} establishes a standardized platform for the development and evaluation of algorithms. By providing a uniform and extensive dataset, complete with \textit{manual annotations} for segmentation and tools for effective 3D visualization, \datasetabbrev{} is designed to develop innovation and enhancement in endovascular intervention. Additionally, the inclusion of \textit{video-based bi-planar} fluoroscopic data enables the exploration of temporal dynamics, such as employing optical flow networks~\cite{nguyen2020end}. \datasetabbrev{} aims to bridge the gap between research innovations and clinical application, targeting essential challenges in endovascular procedures.
\vspace{-2ex}

%% file: sections/1_related_works.tex
\section{Related Work}~\label{sec:related_works}

\textbf{Endovascular Datasets.} Datasets are pivotal for the progression of endovascular navigation, serving as a crucial resource for developing, evaluating, and enhancing algorithms. These datasets, sourced from modalities like mono X-ray~\cite{barbu2007hierarchical,ambrosini2017fully,yi2020automatic,nguyen2020end}, 3D Echo~\cite{wu2014fast}, and 3D MRI~\cite{mastmeyer2017model}, encompass both real and synthetic imagery. However, the limited availability of comprehensive, public datasets for tool segmentation and 3D reconstruction remains a considerable challenge~\cite{barbu2007hierarchical,ambrosini2017fully,mastmeyer2017model,nguyen2020end,danilov2023use}. Table~\ref{tab:dataset-comparison} reveals the predominance of mono X-ray datasets for endovascular interventions, which lack the necessary detail for precise 3D reconstruction in previous works.

\textbf{Catheter and Guidewire Segmentation.}\label{subsec:related-works-segmentation}
The segmentation of endovascular tools, notably guidewires and catheters, represents an advancing field contingent upon dataset availability and quality. Table~\ref{tab:dataset-comparison} illustrates that prior studies have employed synthetic and semi-synthetic data to mitigate the challenges of limited real-world datasets. Works by Barbu~\etal~\cite{barbu2007hierarchical}, Ambrosini~\etal~\cite{ambrosini2017fully}, and Mastmeyer~\etal~\cite{mastmeyer2017model} have utilized manually annotated datasets from 2D X-ray and 3D MRI modalities for training segmentation models. Conversely, Nguyen~\etal~\cite{nguyen2020end} and Danilov~\etal~\cite{danilov2023use} have demonstrated the utility of synthetic datasets in enhancing model efficiency. The introduction of deep learning techniques, particularly U-Net architectures~\cite{ronneberger2015u}, has substantially improved the precision in segmentation and tracking of these surgical instruments. This progress enables fully automated segmentation frameworks that leverage extensively annotated data and employ unsupervised techniques, such as optical flow. Nonetheless, the lack of a public, standardized dataset for method comparison continues to hinder the progression and evaluation of scientific advancements in this area.

\input{tables/endovascular_datasets}

\textbf{3D Reconstruction.}\label{subsec:related-works-reconstruction}
Enhancing 3D reconstruction accuracy in endovascular procedures contributes to superior clinical outcomes by facilitating catheter navigation via advanced visualization and precise tracking. Progress in fluoroscopic imaging technology has enabled more accurate positioning of devices. Wagner~\etal~\cite{wagner20164d} developed an algorithm that leverages elastic grid registration and epipolar geometry for 3D reconstruction from biplane angiography. Hoffmann~\etal~\cite{hoffmann2015electrophysiology,hoffmann2013reconstruction,hoffmann2012semi} employed automatic catheter detection and 3D reconstruction methods utilizing triangulation and graph-search algorithms in electrophysiology studies. Delmas~\etal~\cite{delmas2015three} and Petković~\etal~\cite{petkovic2014real} demonstrated the significance of accurate 3D models for navigation in both complex and single-view vascular architectures, emphasizing the utility of biplanar data. However, the scarcity of comprehensive, publicly accessible datasets for algorithm development and validation in 3D reconstruction represents a significant barrier to both technological advancement and clinical application. This situation highlights the essential need for specialized datasets to foster continued innovation in endovascular tool reconstruction.

%% file: tables/endovascular_datasets.tex
\begin{table}[t]
\renewcommand{\arraystretch}{0.7}
\centering
\resizebox{\textwidth}{!}{
\begin{tabular}{l@{\hskip 0.2in} l l r @{\hskip 0.1in} l l l c}
\toprule
\thead{Dataset} & \thead{Data Collection}  & \thead{Data \\ Type}  & \thead{\#Frames} & \thead{Data \\ Source} & \thead{Annotation} & \thead{Public} & \thead{Task} \\ 
\midrule
Barbu et al.~\cite{barbu2007hierarchical}  
& Mono X-ray & Video & 535  & Real & Manual & No & \multirowcell{8}{Segmentation} \\ \cmidrule[.001mm]{1-7}
Wu et al.~\cite{wu2014fast}             
& 3D Echo & Video & 800  & Real & Manual & No \\  \cmidrule[.001mm]{1-7}
Ambrosini et al.~\cite{ambrosini2017fully} 
& Mono X-ray & Image  & 948  & Real & Manual & No  \\  \cmidrule[.001mm]{1-7}
Mastmeyer et al.~\cite{mastmeyer2017model}  
& 3D MRI & Image  & 101 & Real & Manual & No  \\  \cmidrule[.001mm]{1-7}
Yi et al.~\cite{yi2020automatic} 
& Mono X-ray & Image  & 2,540 & Synthesis & Automatic & No \\  \cmidrule[.001mm]{1-7}
Nguyen et al.~\cite{nguyen2020end} 
& Mono X-ray & Image  & 25,271  & Phantom & Semi-Auto & No  \\  \cmidrule[.001mm]{1-7}
Danilov et al.~\cite{danilov2023use} 
& 3D Ultrasound & Video  & 225 & Synthetic & Manual & No  \\  

\cmidrule{1-8}

Wagner et al.~\cite{wagner20164d}     
& Mono X-ray & Video & - & \makecell[l]{Phantom} & Automatic & No & \multirowcell{9}{3D Reconstruction} \\ \cmidrule[.001mm]{1-7}
Hoffmann et al.~\cite{hoffmann2015electrophysiology} 
& Mono X-ray & Image & 176 & \makecell[l]{Phantom} & Semi-Auto & No  \\  \cmidrule[.001mm]{1-7}
Delmas et al.~\cite{delmas2015three} 
& Mono X-ray & Image  & \makecell[r]{2,289 \\ 5 \\ 63} & \makecell[l]{Simulated \\ Phantom \\ Clinical} & Automatic & No \\  \cmidrule[.001mm]{1-7}
Baur et al.~\cite{baur2016automatic} 
& Bi-planar X-ray & Image  & 70 & Canine & Manual & No \\ \cmidrule[.001mm]{1-7}
Brost et al.~\cite{brost2010catheter} 
& Mono X-ray & Image  & 938 & Clinical & Semi-Auto & No \\  \cmidrule[.001mm]{1-7}
Ma et al.~\cite{ma2010real} 
& Mono X-ray, CT & Image & 1,048 & Clinical & Manual & No \\  \cmidrule[.001mm]{1-7}
Hoffman et al.~\cite{hoffmann2012semi} 
& Bi-planar X-Ray& Video  & 33 &  Clinical & Semi-Auto & No \\  
\cmidrule{1-8}
\datasetabbrev{} (ours) 
& Bi-planar X-ray & \makecell{Video} & 8,746 & Phantom & Manual & Yes & \makecell{ Segmentation \\ 3D Reconstruction} \\ 
\bottomrule
\end{tabular}}
\vspace{1ex}
\caption{Endovascular intervention datasets comparison.}
\label{tab:dataset-comparison}
\vspace{-7ex}
\end{table}

%% file: sections/2_the_dataset.tex
\section{The \datasetabbrev{} Dataset}\label{sec:dataset}
\subsection{Data Collection Setup}~\label{subsec:materials}
\begin{figure}[t]
    \centering 
    \begin{minipage}{0.7\textwidth}
    \subfloat[]{\includegraphics[width=\linewidth]{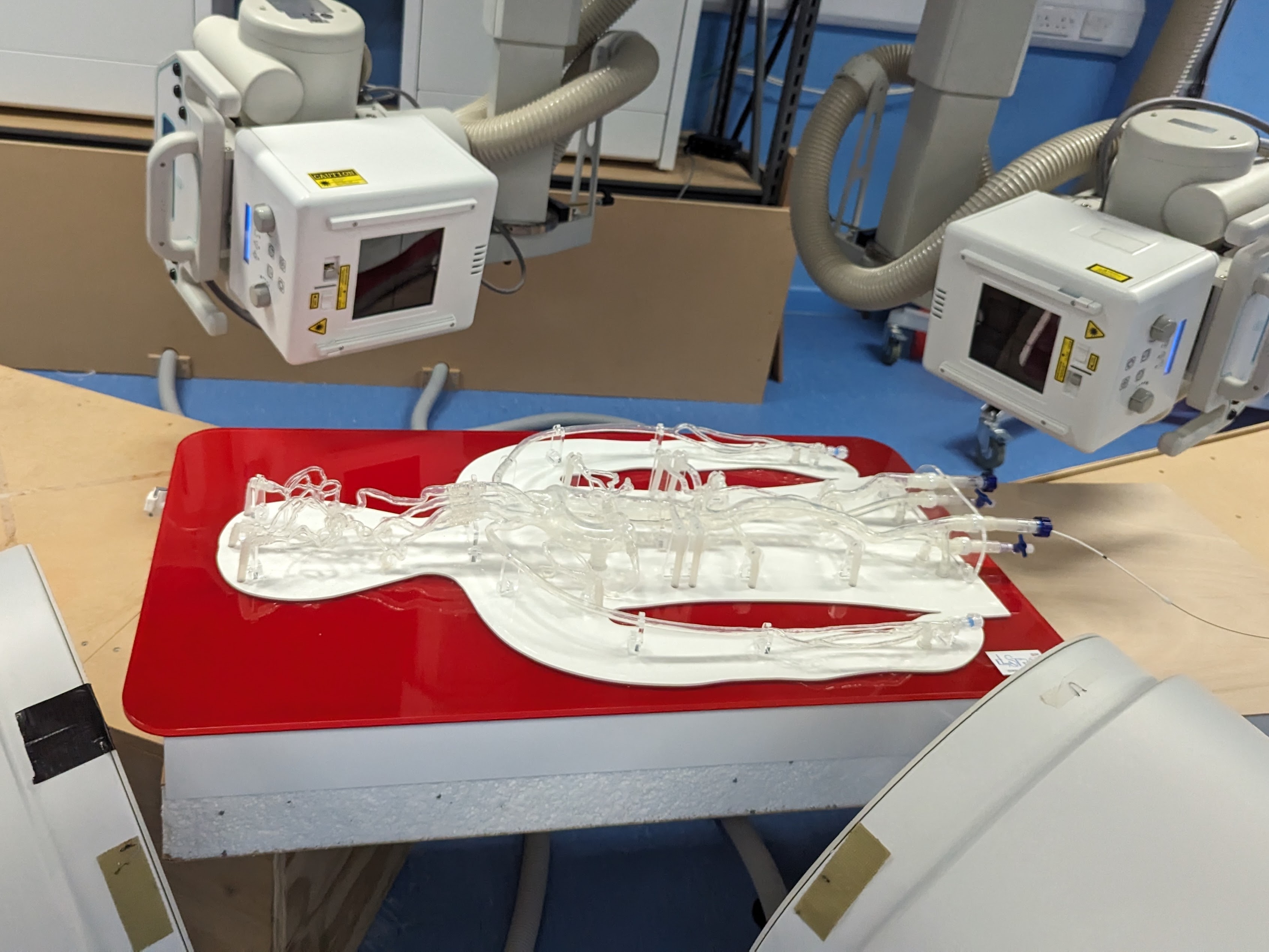}\label{fig:setup}} \\ 
    \end{minipage}\hfill%
    \begin{minipage}{0.25\textwidth}
    \subfloat[]{\includegraphics[width=.99\linewidth]{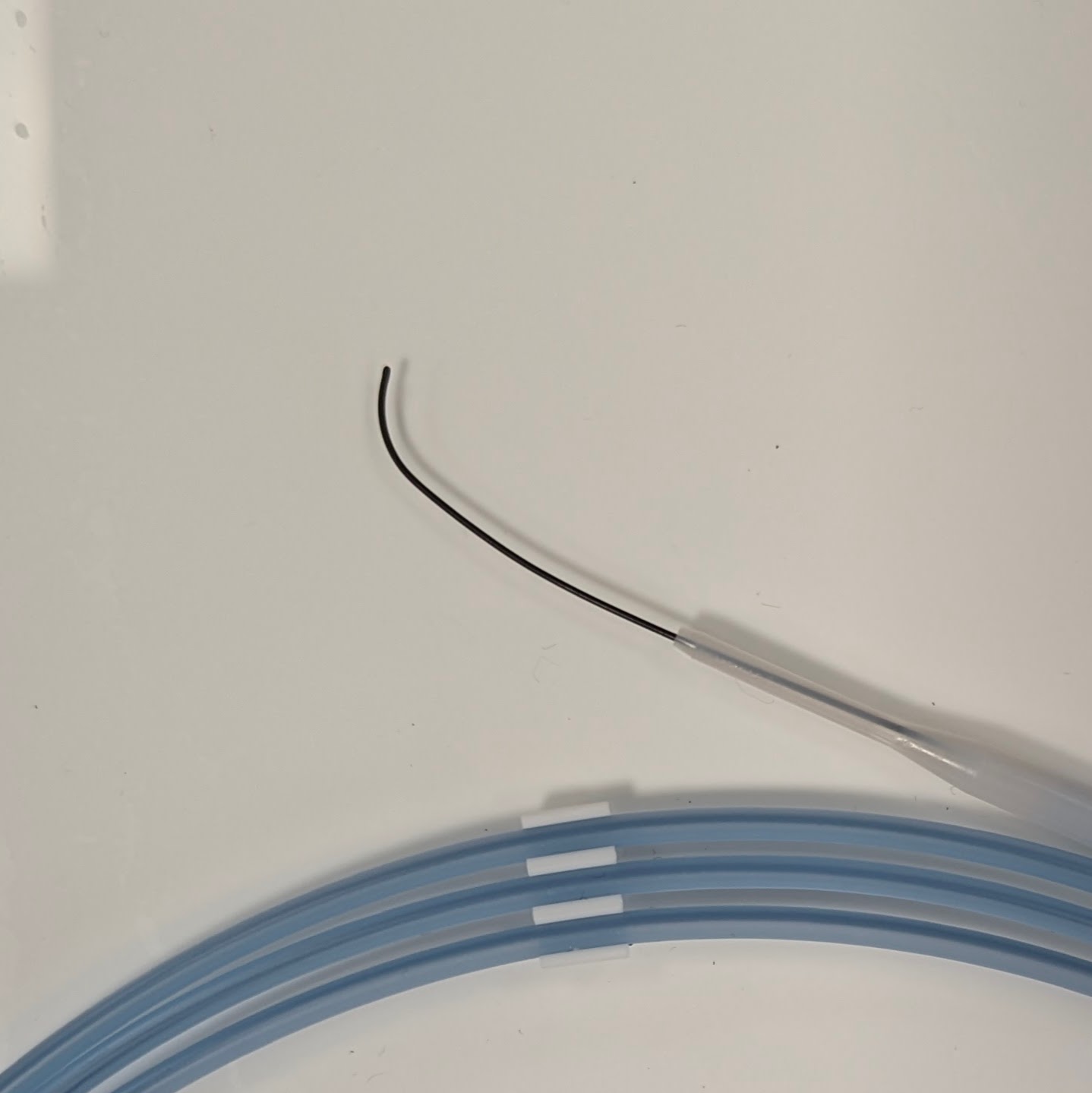}\label{fig:guidewire-nitrex}}  \\
    \subfloat[]{\includegraphics[width=.99\linewidth]{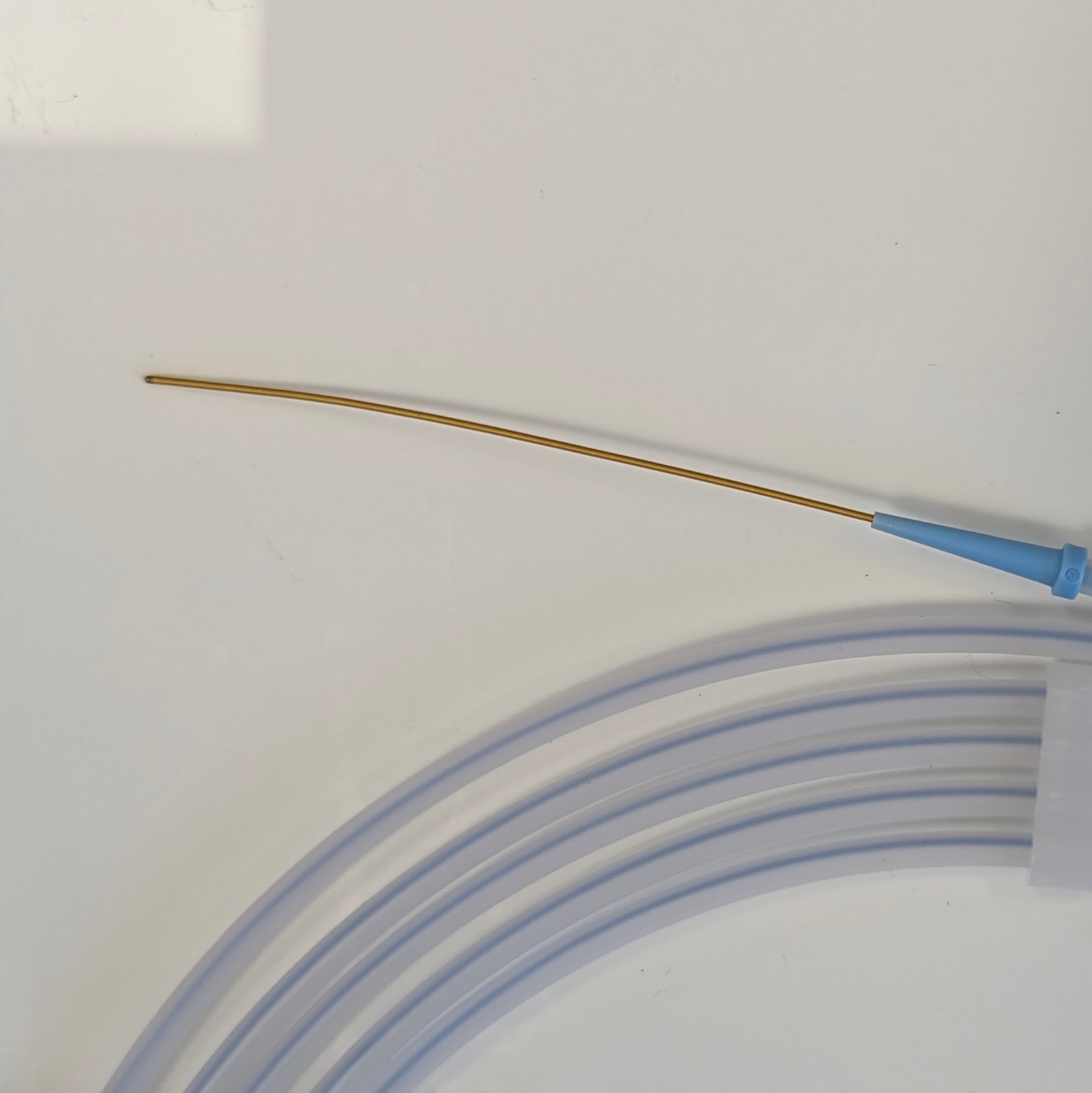}\label{fig:guidewire-radifocus}}
    \end{minipage}
    \caption{\textbf{Materials:} \textit{a)} Overall setup \& endovascular phantom, \textit{b)} Radifocus (angled) guidewire. and \textit{c)} Nitrex (straight) guidewire.}
    \vspace{-1em}
\end{figure}

\textbf{X-ray System.} Our setup utilized a Bi-planar X-ray system (Fig.~\ref{fig:setup}) with \(60\unit{\kilo\watt}\) Epsilon X-ray Generators from EMD Technologies Ltd. and \(16\)-inch Image Intensifier Tubes by Thales, featuring dual focal spot Varian X-ray tubes for high-definition imaging. The setup included Ralco Automatic Collimators for precise alignment and exposure, with calibration achieved through acrylic mirrors and geometric alignment grids.

\textbf{Anatomical Models.}\label{subsubsec:anatomical-models}
Our study utilizes the half-body vascular phantom model from Elastrat Sarl Ltd., Switzerland (Fig.~\ref{fig:setup}). This model is enclosed in a transparent box and integrated into a closed water circuit to simulate blood flow. Constructed from soft silicone and incorporating compact continuous flow pumps with a slippery liquid, it replicates human blood flow dynamics. The design is based on detailed postmortem vascular casts, ensuring anatomical accuracy reflective of human vasculature as documented in seminal works~\cite{martin1998vitro,gailloud1999vitro}. Our utilization of this anatomically accurate model facilitates the realistic simulation of vascular scenarios.

\textbf{Surgical Tools.}\label{subsubsec:surgical-tools}
We enhance our dataset by navigating complex vascular structures using two distinct types of guidewires that are widely used in real-world endovascular surgery. The first, the Radifocus™ Guide Wire M Stiff Type (Terumo Ltd.) (Fig.~\ref{fig:guidewire-radifocus}), is made from nitinol and coated with polyurethane-tungsten. It has a diameter of \(0.89 \unit{\milli\meter}\) and a length of \(260 \unit{\centi\meter}\), featuring a \(3 \unit{\centi\meter}\) angled tip. This guidewire is designed for seeking, dissecting, and crossing lesions. The second, the Nitrex Guidewire (Nitrex Metal Inc.) (Fig.~\ref{fig:guidewire-nitrex}), also made of nitinol, has a gold-tungsten straight tip for enhanced radiopacity within fluoroscopic visualization. It measures \(0.89 \unit{\milli\meter}\) in diameter and \(400 \unit{\centi\meter}\) in length with a \(15 \unit{\centi\meter}\) tip. Unlike the Radifocus guidewire, the Nitrex guidewire is generally used for accessing or maintaining position while exchanging catheters. Both guidewires were selected for their relevance to the operation in real-world settings and to maximize the diversity of the data in our dataset.

\input{tables/descriptive}

\subsection{Data Acquisition, Labeling, and Statistics}\label{subsec:acquisition-labelling-statistics}
Utilizing the materials described in Subsection~\ref{subsec:materials}, we compiled a dataset of 8,746 high-resolution samples (\(1,024 \times 1,024\) pixels). This dataset includes 4,373 paired instances with and without a simulated blood flow medium. Specifically, it comprises 6,136 samples from the Radifocus Guidewire and 2,610 from the Nitrex guidewire, establishing a robust foundation for automated guidewire tracking in bi-planar scanner images. Manual annotation was performed using the Computer Vision Annotation Tool (CVAT)~\cite{cvat2023}, where polylines were created to accurately track the dynamic path of the guidewire. The decision to represent the guidewire as a polyline was due to the inherent structure of a guidewire, where certain parts would inevitably overlap, making a segmentation mask an unsuitable representation. In contrast, a polyline can effectively represent a looping guidewire, providing better accuracy. As detailed in Table~\ref{tab:dataset-stats}, the dataset includes 3,664 instances of angled guidewires with fluid and 484 without, while straight guidewires are represented by 2,472 instances with fluid and 2,126 without. This distribution illustrates a variety of procedural contexts. We note that all 8,746 images in our dataset are equipped with \textit{manual segmentation ground truth}, hence supporting the development of algorithms that need segmentation maps as the reference.

\subsection{Calibration}~\label{subsec:calibration}
\begin{figure}[t]
    \centering
    \subfloat[undistortion]{\includegraphics[width=.48\textwidth]{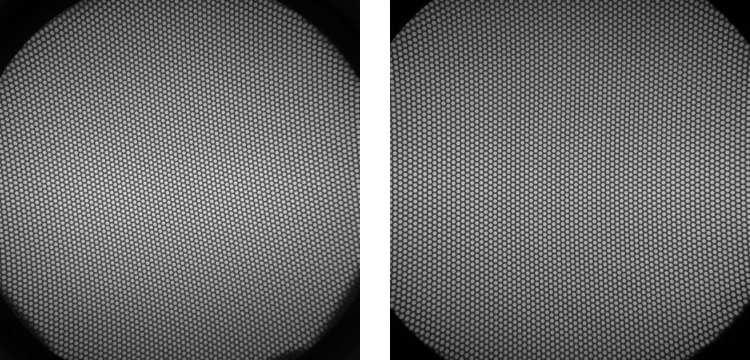}}\hfill
    \subfloat[calibration]{\includegraphics[width=.48\textwidth]{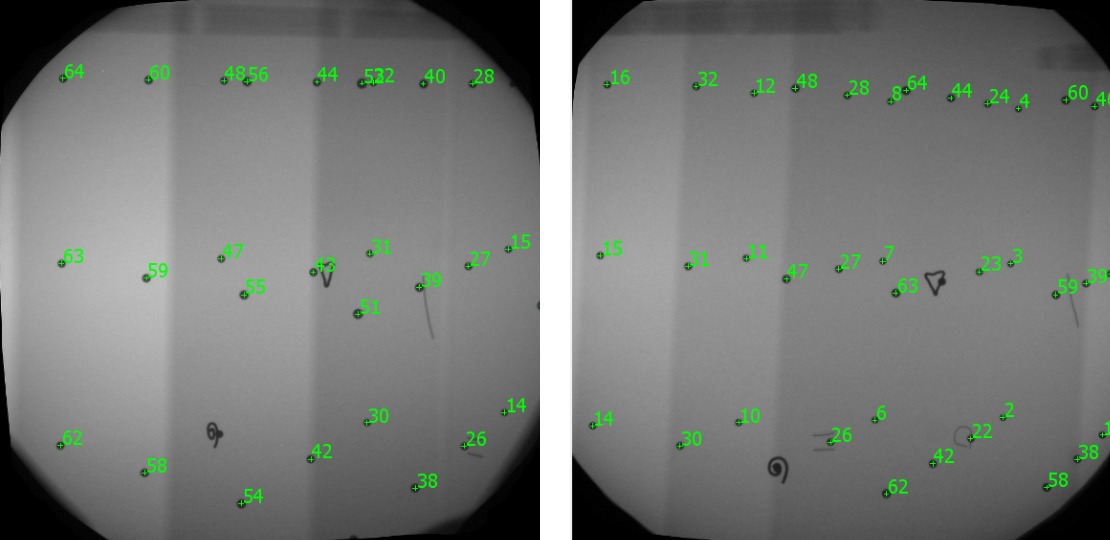}}
    \caption{\textbf{Fluoroscopic Calibration:} \textit{a)} Undistortion grid application, and \textit{b)} Point identification on calibration frame.}
    \label{fig:calibration}
    \vspace{-5pt}
\end{figure}

We extract the camera parameters by following the traditional undistortion and calibration method. Initially, undistortion is achieved using a local weighted mean (LWM) algorithm, applying a perforated steel sheet with a hexagonal pattern (from McMaster-Carr Ltd., part number 9255T645) as framing reference, and employing a blob detection algorithm for precise identification of distortion points~\cite{brainerd2010x}. This setup establishes correspondences between distorted and undistorted positions, enabling accurate distortion correction as per Verdonck~\etal~\cite{verdonck1999variations}. A subsequent calibration step is undertaken utilizing a semi-automatic approach for marker identification and a random sampling consensus (RANSAC) method to ensure robustness in computing the projection matrix and deriving intrinsic and extrinsic camera parameters. The calibration process is refined through direct linear transformation (DLT) and non-linear optimization, considering multiple poses of the calibration object to optimize the entire camera setup~\cite{zhang2000flexible}. Figure~\ref{fig:calibration} shows the calibration process.

\subsection{Utility of \datasetabbrev{} Dataset for the Research Community}\label{subsec:utility}
\datasetabbrev{} advances endovascular imaging by offering a bi-planar fluoroscopic dataset for \textit{segmentation} and \textit{3D reconstruction}, serving as an open-source benchmark. It supports precise algorithm comparisons for segmentation~\cite{ambrosini2017fully,subramanian2019automated,nguyen2020end,zhou2020real} and facilitates method development in 3D reconstruction with its use of bi-planar imagery~\cite{ambrosini20153d,baur2016automatic,brost2010catheter,ma2010real}. With video data,~\datasetabbrev{}{} enables \textit{video}-based methods, leveraging temporal dimensions for dynamic analysis, which enriches the segmentation and reconstruction capabilities whilst also adhering to the nature of the procedure. This versatility underscores \datasetabbrev{}'s crucial role in advancing endovascular imaging~\cite{ramadani2022survey}.

%% file: tables/descriptive.tex
\begin{table}[t]
\centering
\setlength{\tabcolsep}{12pt}

\begin{tabular}{c 
    S[table-number-alignment = center] 
    S[table-number-alignment = center] 
    S[table-number-alignment = center] 
}
\toprule
\thead{Sample Type} & {\thead{Radifocus™ Guide Wire \\ (Angled)}} & {\thead{Nitrex Guidewire \\ (Straight)}} & {\thead{Total}} \\
\midrule
w fluid   & 3664 & 484   & 4148 \\
w/o fluid & 2472 & 2126 & 4598 \\
\cmidrule{1-4}
Total & 6136 & 2610 & 8746 \\
\bottomrule
\end{tabular}
\vspace{1em}
\caption{Dataset Composition Overview.}\label{tab:dataset-stats}
\vspace{-20pt}
\end{table}

%% file: sections/3_shape_prediction.tex
\section{Guidewire Shape Prediction}\label{sec:3D-shape-prediction}
Utilizing the Guide3D dataset, in this section, we build a benchmark for the shape prediction task, which is a popular task in endovascular intervention~\cite{nguyen2020end}. Accurate shape prediction of the guidewire is essential for successful navigation and intervention. In particular, we present a novel shape prediction network aimed at predicting the guidewire shape from a sequence of monoplanar images. The motivation behind this approach lies in the potential of deep learning to learn spatio-temporal correlations from a static camera with a dynamic scene~\cite{sitzmann2019deepvoxels}. Unlike the conventional reconstruction methods requiring biplanar images~\cite{burgner2011toward}, our network leverages a sequence of images to provide temporal information, enabling the network to learn a mapping from a single image \( \mathbf{I}_A \) to the 3D guidewire curve \( \mathbf{C}(\mathbf{u}) \). By adopting deep learning techniques, we aim to simplify the shape prediction process while maintaining high accuracy. This method holds promise for enhancing endovascular navigation by providing real-time, accurate predictions of the guidewire shape, ultimately improving procedural outcomes and reducing reliance on specialized equipment.

\subsection{Spherical Coordinates Representation}\label{subsec:spherical}
\begin{figure}[t]
    \centering
    \subfloat[Spherical Representation]{\includegraphics[width=.25\textwidth]{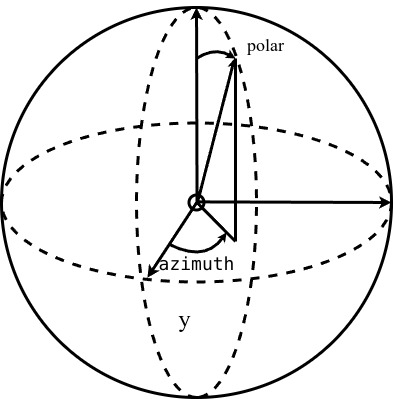}\label{fig:spherical-representation}}\hfill
    \subfloat[Network Architecture]{\includegraphics[width=.7\textwidth]{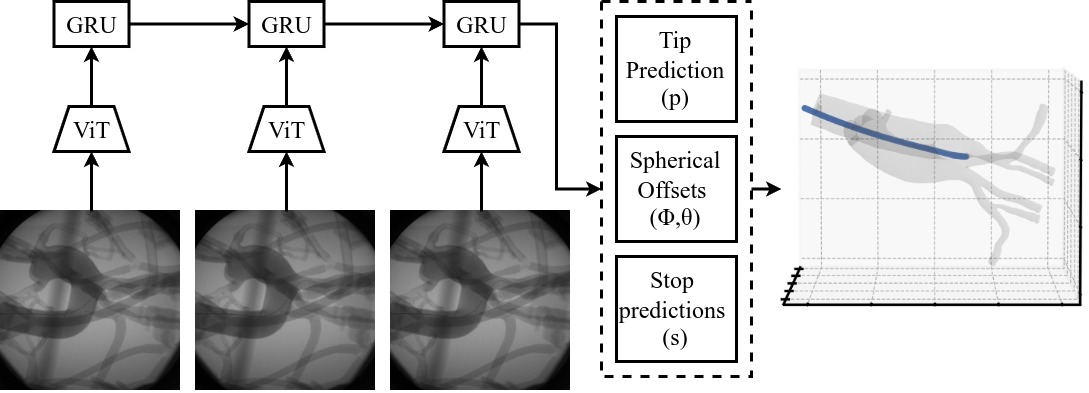}\label{fig:network-architecture}}
    \caption{\textbf{Network Key Components:} The figure illustrates essential components of the proposed model. \textit{a)} Spherical coordinates \((r, \theta, \phi)\) used for predicting the guidewire shape. \textit{b)} The model predicts the 3D shape of a guidewire from image sequences \(\mathbf{I}_t\). A Vision Transformer (ViT) extracts spatial features \(\mathbf{z}_t\), which a Gated Recurrent Unit (GRU) processes to capture temporal dependencies, producing hidden states \(\mathbf{h}_t\). The final hidden state drives three prediction heads: the Tip Prediction Head for 3D tip position \(\mathbf{p} \in \mathbb{R}^3\), the Spherical Offset Prediction Head for coordinate offsets \((\Delta\phi, \Delta\theta)\), and the Stop Prediction Head for terminal point probability \(\mathbf{S}\).}
    \vspace{-1em}
\end{figure}

Predicting 3D points directly can be challenging due to the high degree of freedom. To mitigate this, we use spherical coordinates, which offer significant advantages over Cartesian coordinates for guidewire shape prediction. Spherical coordinates, as represented in Fig.~\ref{fig:spherical-representation}, defined by the radius \( r \), polar angle \( \theta \), and azimuthal angle \( \phi \), provide a more natural representation for the position and orientation of points along the guidewire, which is typically elongated and curved. Mathematically, a point in spherical coordinates \((r, \theta, \phi)\) can be converted to Cartesian coordinates \((x, y, z)\) using the transformations \( x = r \sin \theta \cos \phi \), \( y = r \sin \theta \sin \phi \), and \( z = r \cos \theta \). This conversion simplifies the modeling of angular displacements and rotations, as spherical coordinates directly encode directional information.

Predicting angular displacements \((\Delta\theta, \Delta\phi)\) relative to a known radius \(r\) aligns with the physical constraints of the guidewire, facilitating more accurate and interpretable shape predictions. By predicting an initial point (tip position) and representing subsequent points as offsets in \(\Delta \phi\) and \(\Delta \theta\) while keeping \(r\) fixed, this method simplifies shape comparison and reduces the parameter space. This approach enhances the model's ability to capture the guidewire's spatial configuration and improves overall prediction performance.

\subsection{Network Architecture}\label{subsec:network-architecture}

The proposed model (shown in Fig.~\ref{fig:network-architecture}) addresses the problem of predicting the 3D shape of a guidewire from a sequence of images. Each image sequence captures the guidewire from different time steps (\(\mathbf{I}_{A,t}\)), and the goal is to infer the continuous 3D shape \(\mathbf{C}_t(\mathbf{u}_t)\). This many-to-one prediction task is akin to generating a variable-length sequence from variable-length input sequences, a technique commonly utilized in fields such as machine translation~\cite{sutskever2014sequence} and video analysis.

To achieve this, the input pipeline consists of a sequence of images depicting the guidewire. A Vision Transformer (ViT)~\cite{dosovitskiy2020image}, pre-trained on ImageNet, is employed to extract high-dimensional spatial feature representations from these images. The ViT generates feature maps \(\mathbf{z}_t \in \mathbb{R}\). These feature maps are then fed into a Gated Recurrent Unit (GRU) to capture the temporal dependencies across the image sequence. The GRU processes the feature maps \(\mathbf{z}_t\) from consecutive time steps, producing a sequence of hidden states \(\mathbf{h}_t\). Formally, the GRU operation at time step \( t \) is defined as:
\begin{equation}
\mathbf{h}_t = \text{GRU}(\mathbf{z}_t, \mathbf{h}_{t-1}),
\end{equation}

The final hidden state \(\mathbf{h}_t\) from the GRU is used by three distinct prediction heads, each tailored for a specific aspect of the guidewire shape prediction:
\begin{itemize}
    \item \textbf{Tip Prediction Head:} responsible for predicting the 3D coordinates of the guidewire's tip. A fully connected layer maps the hidden state \(\mathbf{h}_t\) to a Cartesian anchoring point \(\mathbf{p} \in \mathbb{R}^3\).
    \item \textbf{Spherical Offset Prediction Head:} predicts the spherical coordinate offsets \((\Delta\phi, \Delta\theta)\) for points along the guidewire with a fixed radius \( r \).
    \item \textbf{Stop Prediction Head:} outputs the probability distribution indicating the terminal point of the guidewire. It uses a softmax layer to produce a probability tensor \(\mathbf{S}\), where each element \(\mathbf{S}_j\) indicates the probability of the \( j \)-th point being the terminal point.
\end{itemize}

\subsection{Loss Function}\label{subsec:loss-fn}

The custom loss function for training the model combines multiple components to handle the point-wise tip error, variable guidewire length (stop criteria), and tip position predictions. The overall loss function \( \mathcal{L}_{\text{total}} \) is defined as:

\begin{equation}\label{eq:loss_fn}
\begin{split} 
\mathcal{L}_\text{total} = \frac{1}{N} \sum_{i=1}^N \bigg(
& \lambda_\text{tip} \left\| \hat{\mathbf{p}}_i - \mathbf{p}_i \right\|^2 +\\
& \lambda_\text{offset} \big( (\hat{\boldsymbol{\phi}}_i - \boldsymbol{\phi}_i)^2 + (\hat{\boldsymbol{\theta}}_i - \boldsymbol{\theta}_i)^2 \big) +  \\
& \lambda_\text{stop} \big( -\mathbf{s}_i \log(\hat{\mathbf{s}}_i) - (1 - \mathbf{s}_i) \log(1 - \hat{\mathbf{s}}_i) \big) \bigg) 
\end{split}
\end{equation} 
where \( N \) is the number of samples, and \(\lambda_{\text{tip}}\), \(\lambda_{\text{offset}}\), and \(\lambda_{\text{stop}}\) are weights that balance the contributions of each loss component. The tip prediction loss (\(\mathcal{L}_{\text{tip}}\)) uses mean squared error (MSE) to ensure accurate 3D tip coordinates. The spherical offset loss (\(\mathcal{L}_{\text{offset}}\)) also uses MSE to align predicted and ground truth angular offsets, capturing the guidewire's shape. The stop prediction loss (\(\mathcal{L}_{\text{stop}}\)) employs binary cross-entropy (BCE) to accurately predict the guidewire's endpoint.

\subsection{Training Details}\label{subsubsec:training-details} The model was trained end-to-end using the loss from Eq.~\ref{eq:loss_fn}. The NAdam~\cite{dozat2016incorporating} optimizer was used with an initial learning rate of \( 1 \times 10^{-4} \). Additionally, a learning rate scheduler was employed to adjust the learning rate dynamically based on the validation loss. Specifically, the ReduceLROnPlateau scheduler was configured to reduce the learning rate by a factor of \(0.1\) if the validation loss did not improve for 10 epochs.  The model was trained for 400 epochs, with early stopping based on the validation loss to further prevent overfitting. 

%% file: sections/4_experiments.tex
\section{Experiments}\label{sec:experiments}

We evaluate our proposed dataset, Guide3D, through a structured experimental analysis, as follows: \textit{i)} initially, we assess the dataset's validity, focusing on reprojection errors and their distribution across the dataset to understand its accuracy (Section~\ref{subsec:dataset-validation}), \textit{ii)} we then explore the applicability of Guide3D in a 3D reconstruction task (Section~\ref{subsec:shape-prediction-result}), and \textit{iii)} finally, we benchmark several segmentation algorithms to assess their performance on Guide3D, providing insights into the dataset's utility (Section~\ref{subsec:segmentation-results}). 

\begin{figure}[t]
	\centering
	\subfloat{\includegraphics{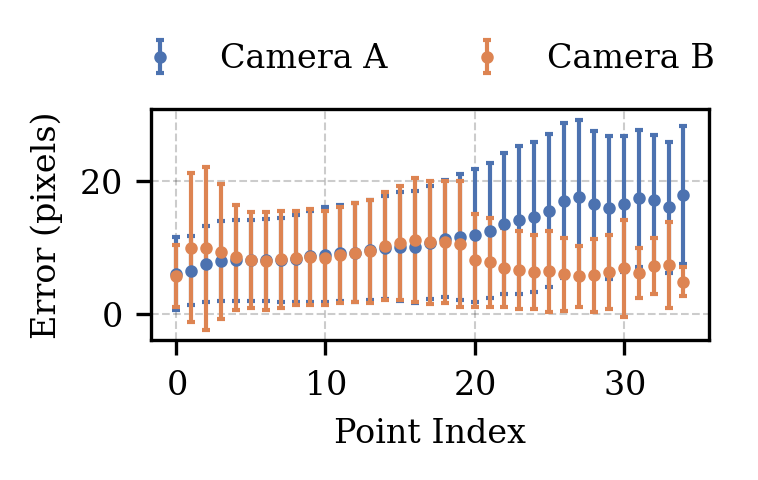}}
	\hfill\raisebox{.3cm}{\subfloat{\includegraphics{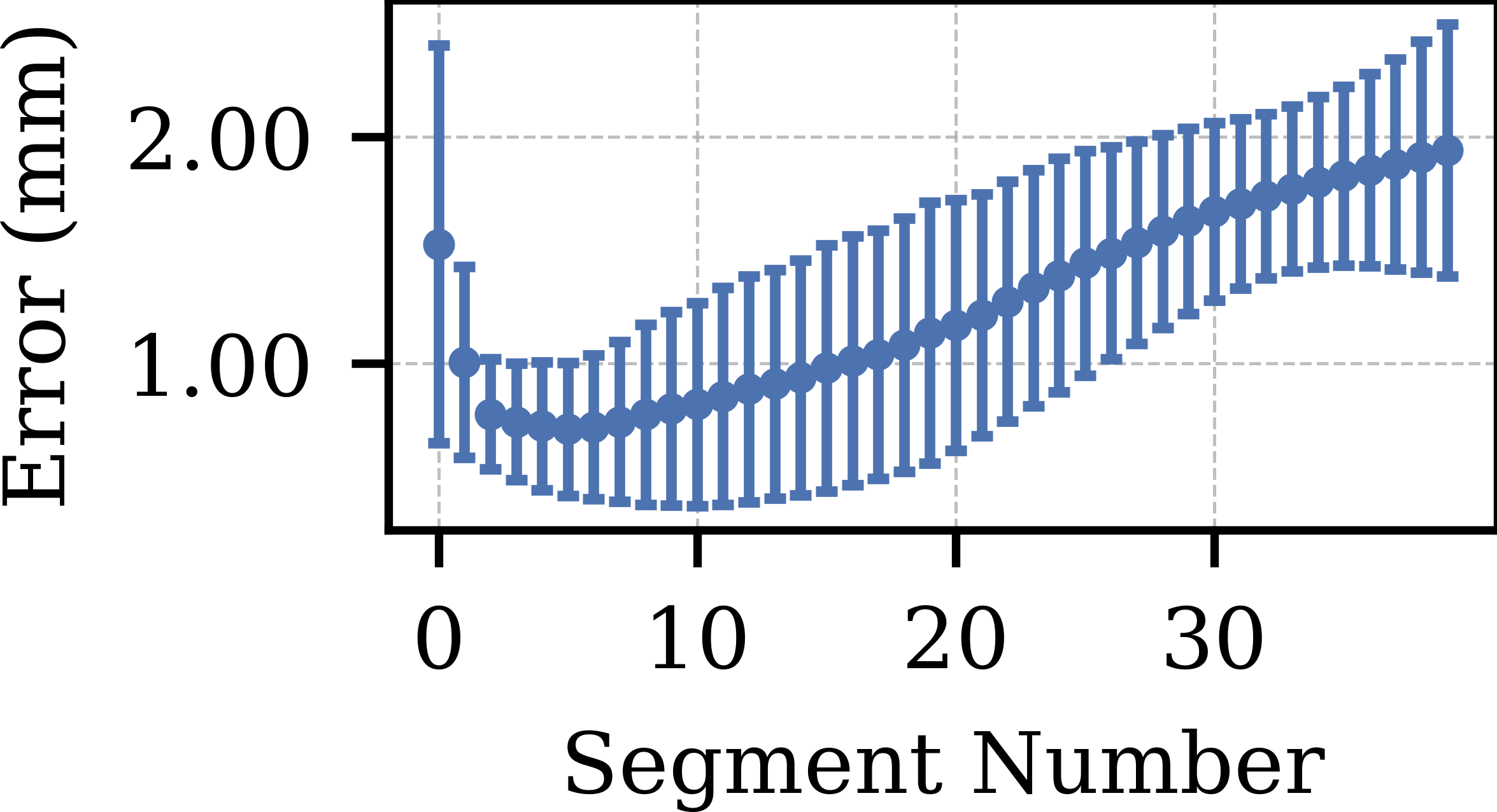}}}
    \caption{\textbf{Guidewire Reconstruction Error Analysis:} (Left) Illustrates the distribution of reprojection errors, noting higher variability and peak errors in the mid-sections and reduced errors at the extremities. (Right) Presents the results of reconstruction validation.}\label{fig:reprojection-error}
\end{figure}

\subsection{Dataset Validation}\label{subsec:dataset-validation}
Our analysis revealed a non-uniform distribution of reprojection errors across the dataset, with the highest variability and errors concentrated at the proximal end of the guidewire reconstructions. Figure~\ref{fig:reprojection-error} shows the reprojection error patterns for both Camera A and Camera B. For Camera A, mean errors increase from approximately \(6\)px to a peak of \(20\)px, with standard deviations rising from \(5\)px to \(11\)px, indicating growing inaccuracies and variability over time. Significant fluctuations around indices \(25\) to \(27\) highlight periods of particularly high error. For Camera B, mean errors exhibit an initial peak of \(9\)px at index 1, followed by fluctuations that decrease towards the end. The standard deviations for Camera B start high at \(11\)px and decrease over time, reflecting a pattern of high initial variability that stabilizes later. These patterns are consistent with the inherent flexibility of the guidewire, which can form complex shapes such as loops.

\begin{minipage}{.9\linewidth}
    \begin{minipage}[t][][t]{0.48\linewidth}
        \begin{minipage}[t]{\textwidth}
            \centering
            \scriptsize
            \input{tables/main_quant}
        \end{minipage}
    \end{minipage}
    \hfill
    \begin{minipage}[t][][t]{0.48\linewidth}
        \centering
        \scriptsize
        \input{tables/segmentation_comparison}
    \end{minipage}
\end{minipage}

Furthermore, we conducted a validation procedure using CathSim~\cite{jianu2024cathsim}, incorporating the aortic arch model described in Subsection \ref{subsec:materials} and a guidewire of similar diameter and properties. For sampling, we employed the soft actor-critic (SAC) algorithm with segmented guidewires and kinematic data, producing realistic validation samples. Evaluation metrics included maximum Euclidean distance (MaxED) at \(2.880 \pm 0.640\) mm, mean error in tip tracking (METE) at \(1.527 \pm 0.877\) mm, and mean error related to the robot's shape (MERS) at \(0.001 \pm 0.000\). These results demonstrate the method's precision.

\subsection{Guidewire Prediction Results}\label{subsec:shape-prediction-result}

We now demonstrate the capability of the network introduced in Sec.~\ref{sec:3D-shape-prediction} and thus highlight the importance of the proposed dataset. We examine the network prediction in the following manner: \textit{1)} we first conduct an analysis between the predicted and reconstructed curve by employing piecewise metrics, and \textit{2)} we showcase the reprojection error.

\begin{figure}[t]
    \centering
    \subfloat[Guidewire Shape Predictions]{\includegraphics[width=.5\textwidth]{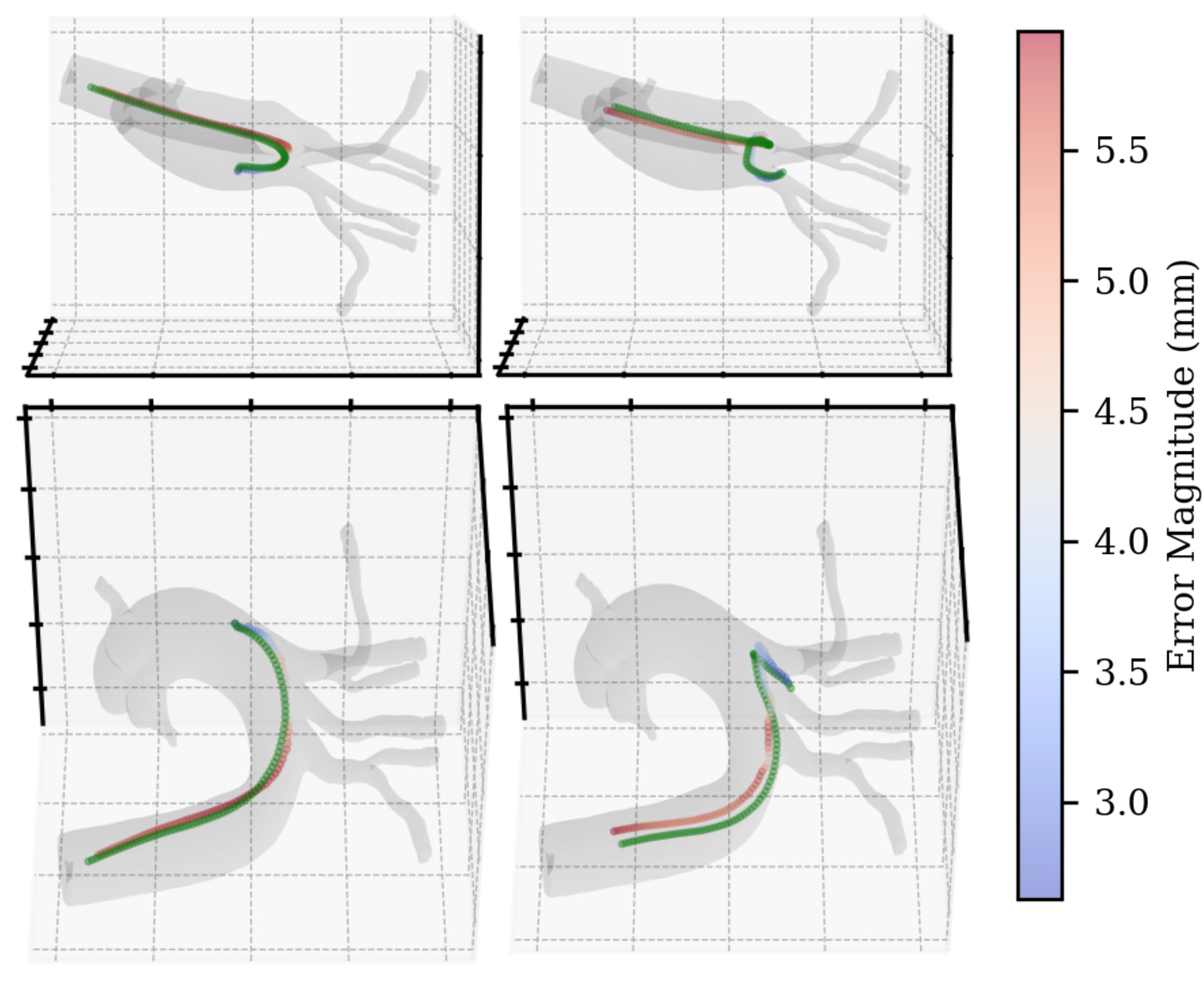}\label{fig:3D}}\hfill
    \subfloat[Reprojections]{\includegraphics[width=.42\textwidth]{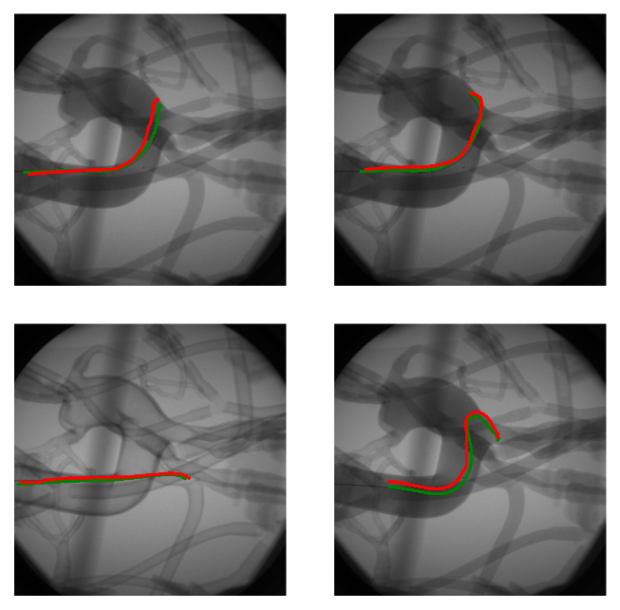}\label{fig:reprojection_visualization}}
    \caption{The figure illustrates the reconstruction similarity of the guidewire when reprojected onto the images. It demonstrates the network's capability to accurately predict the guidewire shape, even in the presence of noticeable angles, highlighting the robustness of the prediction model. }
\end{figure}

\subsubsection{Shape Prediction Errors} Table~\ref{tab:shape-accuracy} presents the comparison of different metrics for shape prediction accuracy. Similarly to Subsection~\ref{subsec:dataset-validation} we quantify the shape differences using the following metrics: \textit{1)} Maximum Euclidean Distance (MaxED), \textit{2)} Mean Error in Tip Tracking (METE), and \textit{3)} Mean Error in Robot Shape (MERS). For all the metrics, the shape of the guidewire, represented as a 3D curve \( \mathbf{C}(u) \), is sampled at equidistant \( \Delta u \) intervals along the arclength parameter \( u \). Therefore, the metrics represent the pointwise discrepancies between the two shapes along the curve's arclength.

The results indicate that the Spherical representation consistently outperforms the Cartesian representation across all metrics. Specifically, the Maximum Euclidean Distance (MaxED) shows a lower error in the Spherical representation (\( 6.88 \pm 5.23 \) mm) compared to the Cartesian representation (\( 10.00 \pm 4.64 \) mm). Similarly, the Mean Error in Tip Tracking (METE) is significantly lower in the Spherical representation (\( 3.28 \pm 2.59 \) mm) than in the Cartesian representation (\( 6.93 \pm 3.94 \) mm). For the Mean Error in Robot Shape (MERS), the Spherical representation also demonstrates a reduced error (\( 4.54 \pm 3.67 \) mm) compared to the Cartesian representation (\( 5.33 \pm 2.73 \) mm). Lastly, the Fréchet distance shows a smaller error for the Spherical representation (\( 6.70 \pm 5.16 \) mm) compared to the Cartesian representation (\( 8.95 \pm 4.37 \) mm). These results highlight the advantage of using the Spherical representation for more accurate shape prediction.

\subsubsection{Shape Comparison Visualization.} Figure~\ref{fig:3D} showcases two 3D plots from different angles, comparing the ground truth guidewire shape to the predicted shape by the network. The network demonstrates its capability to accurately predict the guidewire shape, even in the presence of a loop and self-obstruction in the image. The predicted shape aligns closely with the actual configuration of the guidewire. Notably, the proximal end manifests a more substantial error relative to the nominal error seen at the distal end. Discrepancies from the authentic guidewire shape span from a mere \( 2 \unit{mm} \) at the distal end to a noticeable \( 5 \unit{mm} \) at the proximal end. Impressively, the network evidences its capability to accurately predict the guidewire's shape using only consecutive singular plane images. Subsequently, the 3D points are reprojected onto the original images, as illustrated in Fig.~\ref{fig:reprojection_visualization}.

\subsection{Segmentation Results}\label{subsec:segmentation-results}
We demonstrate \datasetabbrev{}'s potential to advance guidewire segmentation research by evaluating the performance of three state-of-the-art network architectures: UNet~\cite{ronneberger2015u} (learning rate: \( 1 \times 10^{-5} \), 135 epochs), TransUnet~\cite{chen2021transunet} (integrating ResNet50 and Vision Transformer [ViT-B-16], learning rate: 0.01, 199 epochs), and SwinUnet~\cite{cao2022swin} (Swin Transformer architecture, learning rate: 0.01, 299 epochs). Performance metrics included the Dice coefficient (DiceM), mean Intersection over Union (mIoU), and Jaccard index, detailed in Table~\ref{tab:segmentation-results}. The results indicate that UNet achieved a DiceM of 92.25, mIoU of 36.60, and Jaccard index of 86.57. TransUnet outperformed with a DiceM of 95.06, mIoU of 41.20, and Jaccard index of 91.10. SwinUnet recorded a DiceM of 93.73, mIoU of 38.58, and Jaccard index of 88.55. These findings benchmark the dataset's performance and suggest potential for future enhancements. Despite these promising results, the presence of loops and occlusions within the guidewire indicates that polyline prediction could significantly improve task utility.

\begin{figure}[t]
    \centering
    \hfill
    \subfloat[]{\includegraphics[width=.4\columnwidth]{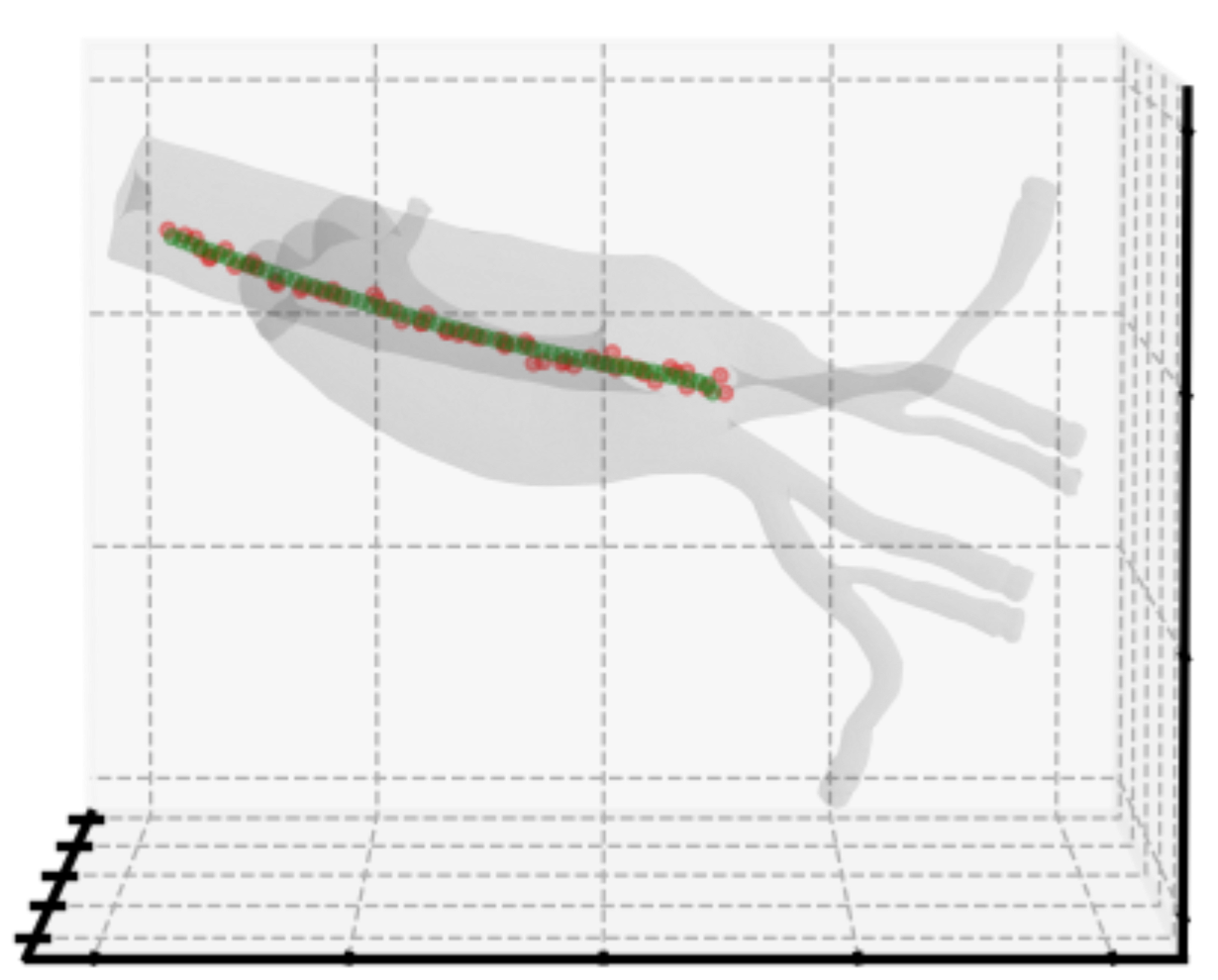}}\hfill
    \subfloat[]{\includegraphics[width=.4\columnwidth]{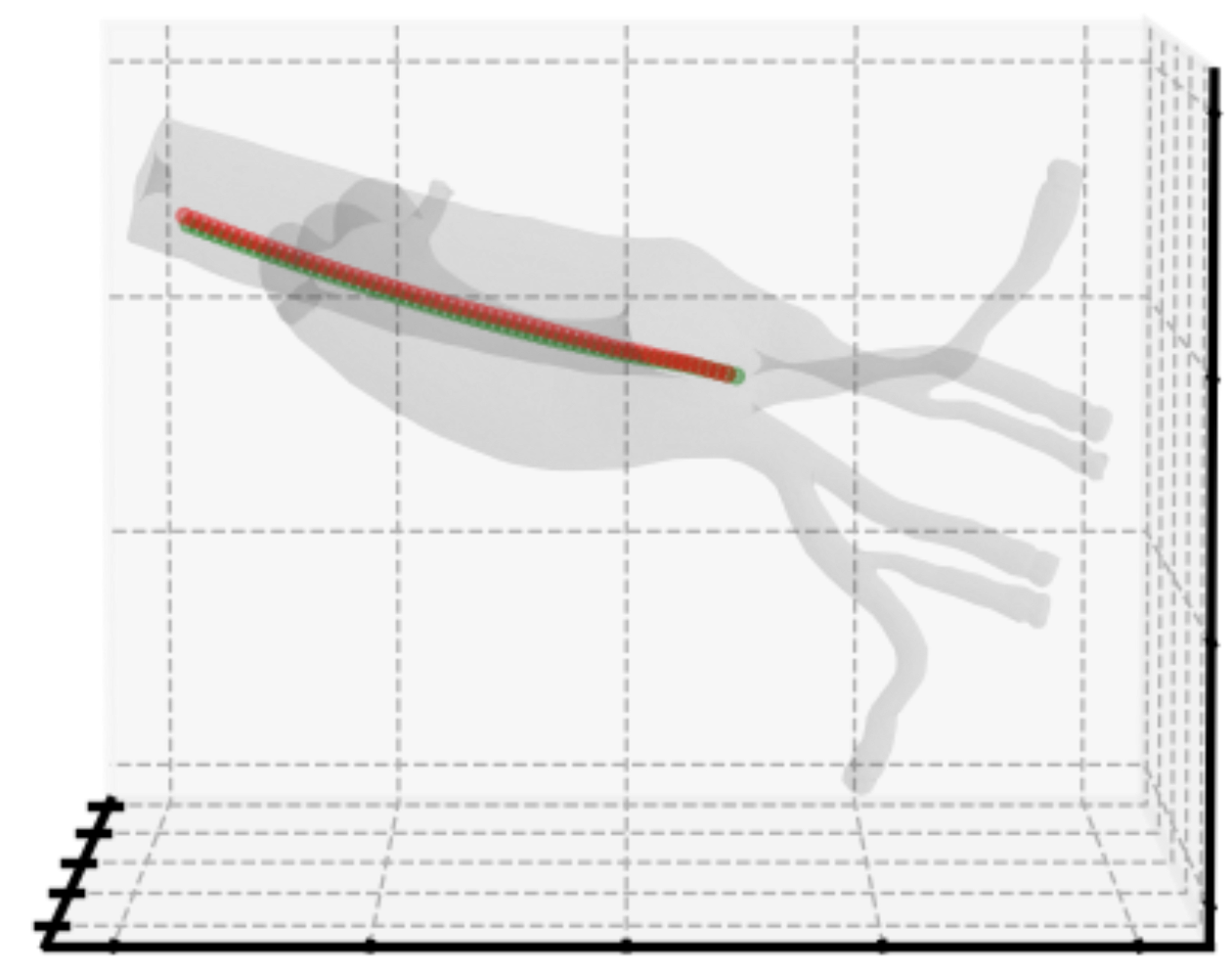}}\hfill
    \caption{Comparison between the predictions of the method not employing spherical coordinates (left) and the method employing spherical coordinates (right). It is evident that the method using spherical coordinates results in a more uniform and accurate shape reconstruction.}
    \vspace{-3ex}
    \label{fig:simple-vs-spherical}
\end{figure}

%% file: tables/main_quant.tex
\begin{tabular}{l r r r r r}
\toprule
&{\hskip 0.15in} \text{\thead{Cartesian}} & {\hskip 0.2in}\text{\thead{Spherical}} \\
\midrule
MaxED~\(\downarrow\) & \(10.00 \pm 4.64\) & \(6.88 \pm 5.23\) \\
METE~\(\downarrow\) & \(6.93 \pm 3.94\) & \(3.28 \pm 2.59\) \\
MERS~\(\downarrow\) & \(5.33 \pm 2.73\) & \(4.54 \pm 3.67\) \\
\bottomrule
\end{tabular}
\captionof{table}{Shape Comparison (mm). }\label{tab:shape-accuracy}

%% file: tables/segmentation_comparison.tex
\begin{tabular}{l 
    S[detect-weight]
    S[detect-weight]
    S[detect-weight]
}
\toprule
 & {\thead{DiceM}} & {\thead{mIoU}} & {\thead{Jaccard}} \\
\midrule
UNet~\cite{ronneberger2015u}      &	92.25  & 36.60  & 86.57 \\
TransUnet~\cite{chen2021transunet} &	 \textbf{95.06} & \textbf{41.20}  & \textbf{91.10}  \\
SwinUnet~\cite{cao2022swin}  & 93.73 & 38.58 &	88.55 \\
\bottomrule
\end{tabular}
\captionof{table}{Segmentation Results}
\label{tab:segmentation-results}

%% file: sections/5_conclusion.tex
\section{Discussion and Conclussion}

In this study, we introduce \datasetabbrev{}, a publicly available bi-planar endovascular navigation dataset designed for the segmentation and 3D reconstruction of flexible, curved endovascular tools, addressing a significant gap in medical imaging research. While extensive experiments demonstrate the dataset's value, we acknowledge limitations such as the absence of clinical real human data due to regulatory challenges and the focus on synthetic and experimental scenarios that may not fully capture the variability of real-world clinical environments. Nevertheless, our standardized platform accommodates both video and image-based approaches, providing a versatile resource to facilitate the translation of these technologies into clinical settings. By including complexities like the guidewire's flexibility and the presence of loops and occlusions, we aim to push the boundaries of current methodologies, although further validation with clinical data is necessary to ensure robustness and generalizability. Our objective is to bridge the disparity between research developments and clinical application by establishing a standardized framework for evaluating various methodologies.